\documentclass{aastex63}

\usepackage{float}
\usepackage{comment}
\usepackage{rotating}

\shorttitle{Improving white dwarf chronometers}
\shortauthors{Moss et al.}

\usepackage{color}
\definecolor{dkgreen}{rgb}{0,0.3,0}

\graphicspath{{./}{figures/}}

\submitjournal{ApJ}

\begin{document}

\title{Improving white dwarfs as chronometers with Gaia parallaxes and spectroscopic metallicities}

\correspondingauthor{Adam Moss}
\email{Adam.G.Moss-1@ou.edu}

\author[0000-0001-7143-0890]{Adam Moss}
\affil{Department of Physics and Astronomy, University of Oklahoma, Norman, OK, USA}

\author[0000-0002-5775-2866]{Ted von Hippel}
\affiliation{Department of Physical Sciences, Embry-Riddle Aeronautical University, Daytona Beach, FL, USA}

\author{Elliot Robinson}
\affiliation{Department of Physical Sciences, Embry-Riddle Aeronautical University, Daytona Beach, FL, USA}

\author[0000-0002-6871-1752]{Kareem El-Badry}
\affiliation{Department of Astronomy, University of California at  Berkeley, Berkeley, CA, USA}

\author{David C. Stenning}
\affiliation{Department of Statistics \& Actuarial Science, Simon Fraser University, Burnaby, BC, Canada}

\author[0000-0002-0816-331X]{David van Dyk}
\affiliation{Department of Mathematics, Imperial College London, London, UK}

\author[0000-0001-9256-5516]{Morgan Fouesneau} 
\affiliation{Max Planck Institute for Astronomy, Heidelberg, Germany}

\author{Coryn A.L.\ Bailer-Jones}
\affiliation{Max Planck Institute for Astronomy, Heidelberg, Germany}

\author{Elizabeth Jeffery}
\affiliation{Physics Department, California Polytechnic State University, San Luis Obispo, CA, USA}

\author{Jimmy Sargent}
\affiliation{Department of Physics and Astronomy, University of Georgia, Athens, GA, USA}

\author{Isabelle Kloc}
\affiliation{Department of Physics, University of Oregon, Eugene, OR, USA}

\author{Natalie Moticska}
\affiliation{Department of Physical Sciences, Embry-Riddle Aeronautical University, Daytona Beach, FL, USA}

\begin{abstract}
White dwarfs (WDs) offer unrealized potential in solving two problems in astrophysics: stellar age accuracy and precision. WD cooling ages can be inferred from  surface temperatures and radii, which can be constrained with precision by high quality photometry and parallaxes. Accurate and precise Gaia parallaxes along with photometric surveys provide information to derive cooling and total ages for vast numbers of WDs. Here we analyse 1372 WDs found in wide binaries with MS companions, and report on the cooling and total age precision attainable in these WD+MS systems.  The total age of a WD can be further constrained if its original metallicity is known, because the main-sequence (MS) progenitor lifetime depends on metallicity at fixed mass, yet metallicity is unavailable via spectroscopy of the WD. We show that incorporating spectroscopic metallicity constraints from 38 wide binary MS companions substantially decreases internal uncertainties in WD total ages compared to a uniform constraint. Averaged over the 38 stars in our sample, the total (internal) age uncertainty improves from 21.04\% to 16.77\% when incorporating the spectroscopic constraint. Higher mass WDs yield better total age precision; for 8 WDs with zero age main sequence masses $\geq$ 2.0 $M_\sun$, the mean uncertainty in total ages improves from 8.61\% to 4.54\% when incorporating spectroscopic metallicities.  We find that it is often possible to achieve 5\% total age precision for WDs with progenitor masses above 2.0 $M_\sun$ if parallaxes with $\leq$ 1\% precision and Pan-STARRS $g, r$, and $i$ photometry with $\leq$ 0.01 mag precision are available.
\end{abstract}

\section{Introduction} \label{sec:intro}

White dwarfs (WDs) are long-established as a means to deriving the age of stellar populations \citep[e.g.][]{Mestel52,Winget87,Liebert88,Knox99,Hansen04,Harris06,vH06,Kilic17}. Because WDs have ceased energy production by fusion, their temperature and luminosity decline monotonically with time, at least in the absence of interactions with a companion. It is therefore often easier to constrain the cooling age or even the total age of a WD than it is to constrain the age of a main sequence (MS) star, the observable properties of which are only weakly sensitive to age. 

Broadly speaking, three parameters of a WD are required in order to estimate its cooling age:  surface temperature, mass, and atmospheric composition.  The surface temperature encodes the WD's current thermal state. The mass constrains both the original thermal energy and core composition of the WD.  The atmospheric composition, typically dominated by hydrogen or helium, dictates the opacity that photons experience as they radiate away the thermal energy of the WD core \citep[see][for further details]{FBB}.  Until recently, the most challenging of these three measurements was mass, which required either high signal-to-noise optical spectroscopy fit with stellar atmosphere models \citep{Bergeron92} or trigonometric parallaxes \citep{Dahn89}.  The spectroscopic technique requires  sufficiently warm photospheres to generate H or He absorption lines, the widths of which are sensitive to atmospheric pressure, which in turn is proportional to surface gravity. The other approach, which can be applied to any WD regardless of surface temperature, is to measure a star's distance via trigonometric parallax, exploiting the fact that distance for a star of a known apparent luminosity and temperature constrains its radius.  Both mass derivation techniques rely on the mass-radius relation for degenerate matter \citep{Hamada61,Vauclair97,Provencal98,Bedard17,Joyce18}.  Previously, the precision on WD masses dominated the cooling and total age uncertainties derived for these stars.  With the release of highly accurate and precise trigonometric parallaxes starting with Gaia DR2 \citep{Gaia18b}, the field of WD research is now able to tease out subtle effects \citep[e.g.][]{Tremblay19} previously obscured by large distance errors.  In addition, Gaia parallaxes allow us to calculate some of the most precise and (occasionally) accurate age estimators \citep{Fouesneau19} determined to date.

Our motivations for improving WDs as chronometers are ultimately focused on deriving an improved star formation history of the Galaxy (to be reported elsewhere), \textbf{refining WD and stellar evolution theory by intercomparing these two systems \citep[e.g.,][]{vH05, DeGennaro09}}, and providing precise ages for binaries that allow additional astrophysics, for instance chemical tagging studies \citep[e.g.][]{Bland-Hawthorn04,Price-Jones20} that have the potential to identify individual star formation episodes. \textbf{In order to accomplish these scientific goals it would be beneficial to improve the precision (and ultimately also the accuracy) of WDs as chronometers.  In this paper, we focus on testing and improving WD age precision by incorporating Gaia trigonometric parallaxes \citep{Gaia18b} and spectroscopic metallicities, the latter of which are from main sequence (MS) companions to WDs in wide binaries.}  We derive the cooling and total ages of 1372 WDs with MS companions. Forty two of these systems have spectroscopic metallicities from the MS components \citep{Cui12, Holtzman18, Steinmetz20}.  We use this subset of 38 WD+MS binaries to demonstrate the additional constraint in total age that results from using the spectroscopic metallicity measurement and its uncertainty for the MS star as a prior on metallicity of the WD.  This constraint is of value both for improving WDs as chronometers and for estimating the expected improvements that come along for free with spectroscopic metallicities measured by large-scale stellar surveys.  \textbf{We note that our paper complements that of \citet{Qiu21}, which also studied a large sample of WDs in binaries with Gaia parallaxes, focusing on validating WD ages within open clusters and WD+WD pairs, as well as deriving the parameters of WD+MS pairs.  The WD+MS binaries we include as well as our photometric and parallax selection criteria are more stringent than \citet{Qiu21} in order for us to perform precision age tests, which are the focus of this paper.}

\section{Selecting WD+MS Binaries} \label{sec:style}

Our sample of WD+MS binaries was derived from the wide binary catalog of \citet{Elbadry18}. This catalog was assembled by searching {\it Gaia} DR2 for pairs of stars within 200 pc of the Sun with positions, parallaxes, and proper motions consistent with bound Keplerian orbits. The catalog contains 3211 high-confidence WD+MS binaries with projected physical separations ranging from a few tens of AU to 50,000 AU and angular separations ranging from a few arcsec to 1 degree. Components of each binary were classified as WD or MS stars based on their location in the color-magnitude diagram. The contamination rate from chance alignments was estimated to be less than 1\%. We refer interested readers to \citet{Elbadry18} and \citet{Elbadry19} for detailed discussions of the wide binary selection procedure, contamination rate, and effective selection function.  Some of these WDs are above the 0.5 $M_\sun$ WD cooling track, suggesting they might be products of binary interactions, or unresolved WD+WD binaries.  These will be discussed further in \S3. 

\textbf{During the course of this study, the Gaia collaboration released early Data Release 3 (eDR3).  We have incorporated eDR3 parallaxes and their uncertainties in the present study.}  The {\it Gaia} bandpasses are suboptimal for constraining WD parameters, both because they are broad and because there are non-negligible uncertainties in the filter response curves \citep[e.g.][]{Weiler18}. We therefore cross-matched the \citet{Elbadry18} WD+MS binary catalog with the Pan-STARRS1 photometric survey \citep{Tonry12, Chambers16}, which obtained {\it grizy} photometry for roughly 3/4 of the sky. We also cross-matched the WDs with the Montreal white dwarf database \citep{Dufour17}, yielding spectral types and effective temperatures for the WDs in 396 binaries (12\% of systems in the \citet{Elbadry18} + Pan-STARRS1 cross-matched catalog). Of the 3211 WD+MS binaries in this cross-matched input catalog, 1396 have a complete set of Pan-STARRS1 $grizy$ photometry and $r$-band precision $\sigma_r \leq 0.2\,\rm mag$ for both components. All the binaries in the catalog have fractional parallax uncertainties of less than 0.05 for the brighter component (i.e. \texttt{parallax\_over\_error} $>$ 20). Additionally, the \citet{Elbadry18} WD+MS binary catalog was cross-matched with a number of wide-field spectroscopic surveys, as described in \citet{Elbadry19b}. This yielded spectroscopic metallicities for the MS star in 38 binaries (four of these will be removed for other reasons in \S3).  \textbf{Line-of-sight absorption values were adopted from the \citet{GF21} catalog, which derived $A_V$ values from 3-D maps of differential extinction (see references therein).  In 98\% of cases our WDs appear in that sample, and we adopt their $A_V$ values as prior means, then set the prior for the standard deviation of $A_V$, $\sigma(A_V)$ = 0.5 $\times A_V$.  The $A_V$ distribution for this WD sample is non-Gaussian, with 91\% of the WDs having $A_V \leq$ 0.1 and the total distribution characterized by mean($A_V$) = 0.047, min($A_V$) = 0.01, and max($A_V$) = 0.96.  For the remaining 2\% of WDs where we have no $A_V$ information, based on the sample mean $A_V$ and assuming a Gaussian distribution of uncertainty on $A_V$ for any given WD, we set the prior mean($A_v$) = 0.05 and $\sigma(A_V)$ = 0.05}. 

With these cuts, the maximum photometric uncertainties among these stars are $\sigma_{g,max}$ = 0.124, $\sigma_{r,max}$ = 0.185, $\sigma_{i,max}$ = 0.107, $\sigma_{z,max}$ = 0.157, and $\sigma_{y,max}$ = 0.400 mag.  Average photometric uncertainties are much less, at ${<}\sigma_g{>}$ = 0.013, ${<}\sigma_r{>}$ = 0.012, ${<}\sigma_i{>}$ = 0.012, ${<}\sigma_z{>}$ = 0.017, and ${<}\sigma_y{>}$ = 0.036 mag.  The average fractional parallax uncertainty, ${<}\sigma_{\varpi} / \varpi{>}$ = 0.011, i.e. just above 1\%.  We require low photometric and parallax uncertainties in order to minimize WD mass uncertainties, which propagate into both the cooling and total age uncertainties.  To study the added benefit of using metallicity as a prior in deriving WD properties, particularly WD total ages, we use the measured metallicities for the MS components from APOGEE \citep{Holtzman18}, LAMOST \citep{Cui12},  and RAVE \citep{Steinmetz20}.  Of the binaries with sufficient photometric precision, 42 have spectroscopic metallicities, though this will be further reduced to 38 (see below).  Our analysis focuses initially on this subset of 38 binaries (Table 1) with metallicity estimates and then expands to the larger group of WD+MS pairs. 
 

\section{Age Constraints Employing Spectroscopic Metallicities} \label{sec:floats}

{\bf B}ayesian {\bf A}nalysis of {\bf S}tellar {\bf E}volution  with {\bf 9} Parameters, abbreviated BASE-9,\footnote{BASE-9 source code is publicly available at \url{https://github.com/BayesianStellarEvolution/base-cpp}.} uses a Markov chain Monte Carlo (MCMC) technique along with optimized numerical integration to estimate the posterior probability distributions for up to nine stellar or cluster parameters: age, mass, distance, metallicity, line-of-site absorption, helium content, binarity, cluster or field membership, and globular cluster subpopulation \citep{vH06,vanDyk09,Stenning16,WK16b}.  \textbf{The first five} parameters on this list are relevant for our present study. We reject BASE-9 fits that would indicate unresolved binaries and the helium content is a fixed value for a specific metallicity in these stellar evolution models.  In this analysis, we employ PARSEC isochrones \citep{Bressan12,Marigo17}, which model stellar phases from the pre-main sequence through the asymptotic giant branch, ages from 26.3 Myr to 13.4 Gyr, and metallicities within the range 0.0005 $<$ Z $<$ 0.07. These models provide synthetic photometry through multiple filter sets, including Gaia and Pan-STARRS. The WD models rely on \citet{WDEC} interiors and \citet{Bergeron95} atmospheres \citep[see also][]{Holberg06}\footnote{\textbf{The models we use were updated by P. Bergeron on 13 Jan 2021.}  See \url{http://www.astro.umontreal.ca/~bergeron/CoolingModels/}}.  We use the Initial-Final Mass Relation (IFMR) of \cite{Cummings2018}.  We employ only a single set of modern models as inputs because our goal is to study the age precision possible with WDs, not the size of systematic uncertainties, which will vary from model to model and likely be a function of mass.  Furthermore, a firm grasp of age precision is an essential stepping stone toward quantifying age systematics among model ingredients, which is beyond the scope of this work.

Hydrogen (DA), helium (DB), and other rarer atmosphere types may yield different ages for the same measured photometry, yet for this sample we know the spectral type for only 12\% of the WDs because of insufficient spectroscopy.  Where a spectral type is unknown, we assume that the star is a DA\footnote{\textbf{Systematic errors introduced by this assumption depend on a WD's mass and cooling age.  Our preliminary study indicates that on average among our sample, assuming a star is a DA rather than a DB decreases the star's age by 0.62 standard deviations of the age posterior.  This issue will be examined further in a future paper.}}, which we expect to be correct for at least three out of every four WDs \citep{GBB19}.  A subset of these WD+MS binaries have measured metallicities. For these we run BASE-9 with the same WD photometry\footnote{We ignore the MS photometry to avoid issues with isochrones potentially being inconsistent with low mass MS photometry.} but with three different prior distributions on metallicity for the MS star.  These three priors are (1) a narrow Gaussian prior base on the \textit{spectroscopic} metallicity and its uncertainty, with a typical value for the uncertainty of 0.075 dex, (2) a wider \textit{nominal} prior that is a Gaussian distribution centered at $-0.2$\,dex with $\sigma=0.5$\,dex, appropriate for a disk WD, and (3) a \textit{uniform} prior distribution over the input model range of [Fe/H] = $-2.0$ to $+0.5$\,dex, appropriate for a WD belonging to an unknown Galactic population.  These three cases test the degree to which the metallicity prior affects the WD cooling age, total age, and mass posterior distributions.  Not all BASE-9 runs converged: 24 WDs among the 1396 could not be fit by the evolution of a single star within the model grid and the age of the Universe.  These stars are likely to be unresolved WD+WD pairs or WDs that resulted from close binary stars with mass transfer.  This left a total of 1372 systems in which BASE-9 derived the WD parameters of cooling age, total age, and Zero Age Main Sequence (ZAMS) mass.  Of the 42 systems that have spectroscopic metallicities for the MS star, four stars were among these 24 systems, and so 38 systems with spectroscopic metallicities remain.  As a note on nomenclature, when we discuss stellar mass, whether for a WD or MS star, we are referring to the ZAMS mass of that star.  This places WD and MS stars on the same system and hopefully avoids confusion because we are using only a single IFMR to map stellar masses from the MS to WD stages.  In addition, when we discuss stellar ages without explicitly writing 'cooling ages' or 'total ages', then either cooling ages are irrelevant (e.g. for MS stars) or the statement applies to both WD cooling ages and total ages.  

\begin{figure}  
    \centering
    \includegraphics[width=1\textwidth]{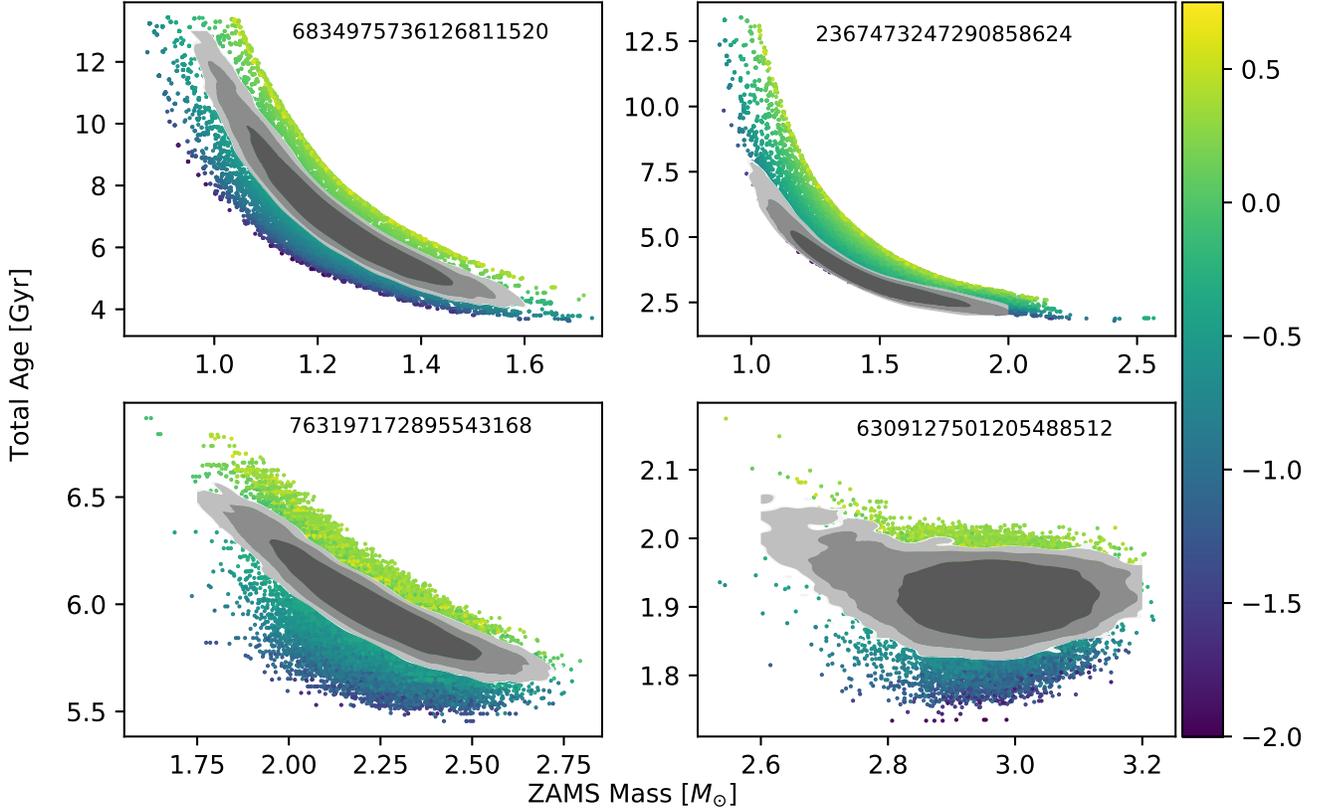}
    \caption{The total age -- ZAMS mass -- metallicity joint posterior probability distributions for four WDs with MS companions with measured spectroscopic metallicities.  The points represent the ZAMS masses, total ages, and metallicities using the nominal metallicity priors with the metallicity of each sample, on the [Fe/H] scale, indicated by the color bar.  The joint posterior distributions on age and mass incorporating the spectroscopic metallicity priors are overplotted as contours. The darkest to lightest contours enclose 34, 68, 95, and 99\% of the posterior probability, respectively.  These contours ignore metallicity, i.e. are summed across all metallicities within these posterior distributions. From upper left through lower right panels the spectroscopic metallicity priors are Gaussian distributions with means (and standard deviations) of [Fe/H] = $-$0.219 (0.101), $-$0.893 (0.17), $-$0.131 (0.05), and $-$0.186 (0.104) dex. These four examples were all derived with H-atmosphere WD models.}
    \label{fig:my_label1}
\end{figure}

Figure 1 presents joint posterior distributions of total age, ZAMS mass, and metallicity for four example systems, chosen to span a range of stellar masses and joint posterior distribution shapes.  These four systems were run with both the nominal and spectroscopic metallicity priors. For the nominal metallicity prior, the figure presents a scatterplot of a sample from joint posterior distribution of age and ZAMS mass, color coded by the sampled values of metallicity. (In this way, the points of a particular color represent a sample from the conditional distribution of age and ZAMS mass given the value of metallicity corresponding to that color.)  Regions of higher point densities are more probable.  The joint probability distributions that incorporate the spectroscopic metallicity priors are overplotted as contours, with light through dark grey contours representing the lowest through highest posterior density. The panels are organized by increasing values of the posterior mean of ZAMS mass. In general, without the spectroscopic metallicity prior, a WD may be consistent with a broader posterior ZAMS mass and/or total age range, although at higher ZAMS mass the metallicity has little impact on the results because of the short evolutionary timescale of high mass stars.

A summary of the cooling ages of all 1372 WDs is presented in Figure 2, which displays the posterior means of the WD cooling ages versus the fractional cooling age uncertainties, defined as one half the width of the 68\% equal-tailed interval divided by the posterior means, e.g. $\sigma_\textrm{68,cooling~age} / <\textrm{cooling~age}>$.  The sample subset of 38 WDs with spectroscopic [Fe/H] estimates for their MS companions are plotted with orange symbols.  Figure 3 presents these 38 systems again, plotted with results computed under their spectroscopic, nominal, and uniform metallicity priors.  The resulting WD cooling ages are highly precise, with most fractional ages uncertainties $\leq$ 0.05.  The cooling ages are independent of the choice of metallicity prior because gravitational settling removes heavy elements from WD photospheres and the energy transport of WD interiors is dominated by the high conductivity of degenerate electrons.  Thus metallicity does not enter as a parameter in WD cooling models.  (Formally, the average ratio of the means of the WD cooling ages calculated under the spectroscopic versus nominal metallicity priors is 1.0002 $\pm$ 0.0055, which is consistent with the WD models being independent of metallicity.)  Figure 4 presents a comparison between WD cooling and total ages.  Objects with metallicity information are dispersed throughout this distribution.  Most of the systems are younger than 6 Gyr due to observational selection, i.e. older WDs are typically too faint to be included in the Gaia sample.  The small excess of objects somewhat younger than 14 Gyr is an indication that some WDs have masses lower than expected from single star evolution and are either unresolved WD+WD binaries or the result of close binary star evolution.  \textbf{These objects would be worth additional follow-up before assigning total system ages with high confidence or for their potential interest as more complicated systems.}

\begin{figure}  
\vspace{0.5cm}
\centering
  \includegraphics[width=0.65\textwidth]{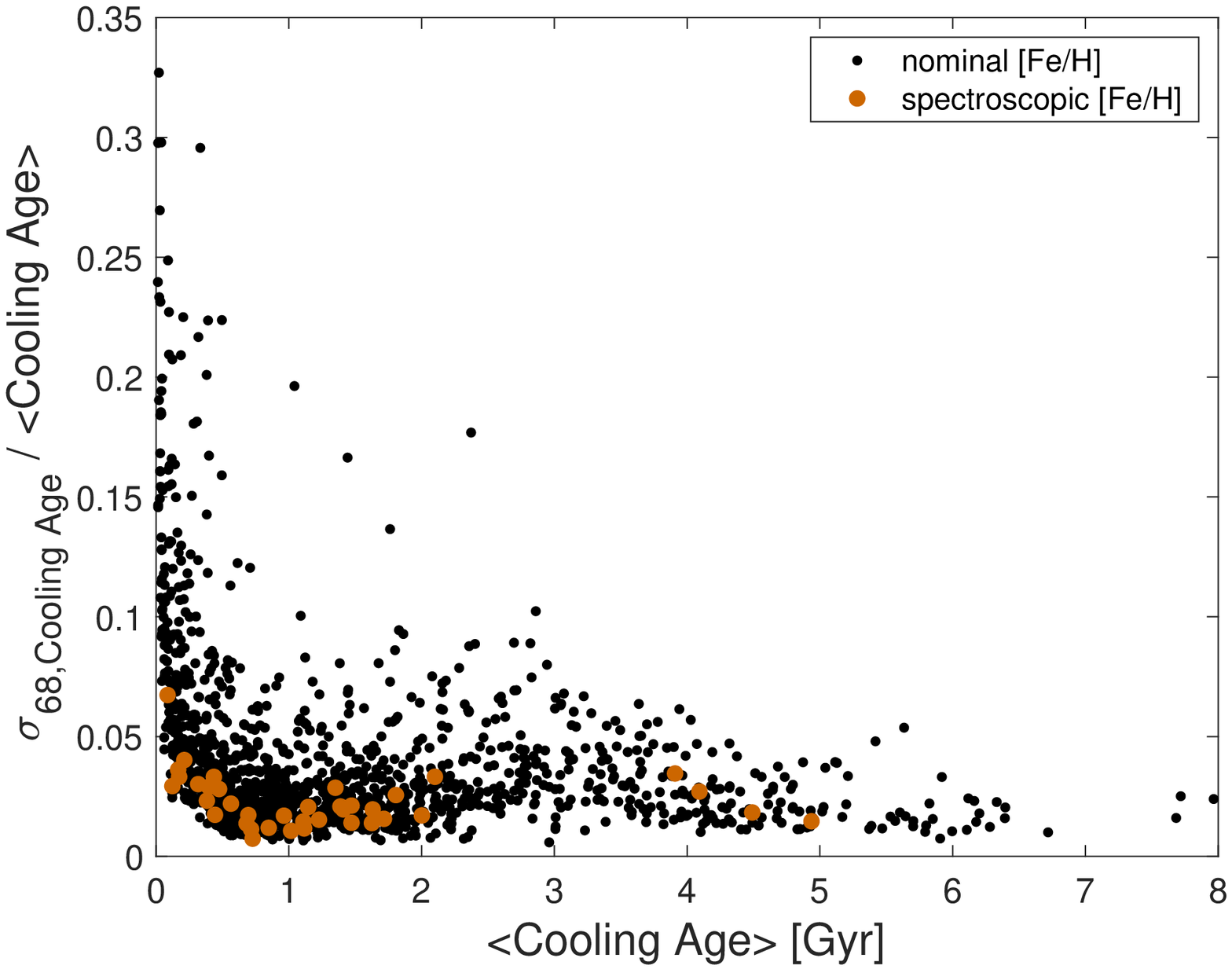}
  \caption{WD cooling age versus fractional uncertainty in cooling ages for 1372 WD+MS binaries.  The uncertainties are defined as one half the width of the 68\% equal-tailed  interval divided by the mean of the posterior distributions of cooling age.  Warmer, more luminous WDs are to the left.  These are over-represented in our sample compared to a volume-limited sample because they are intrinsically brighter and therefore easier to detect.  The subset of objects with spectroscopic metallicities are indicated by the larger orange circles.}
\end{figure}

\begin{figure}  
\centering
  \includegraphics[width=0.45\textwidth]{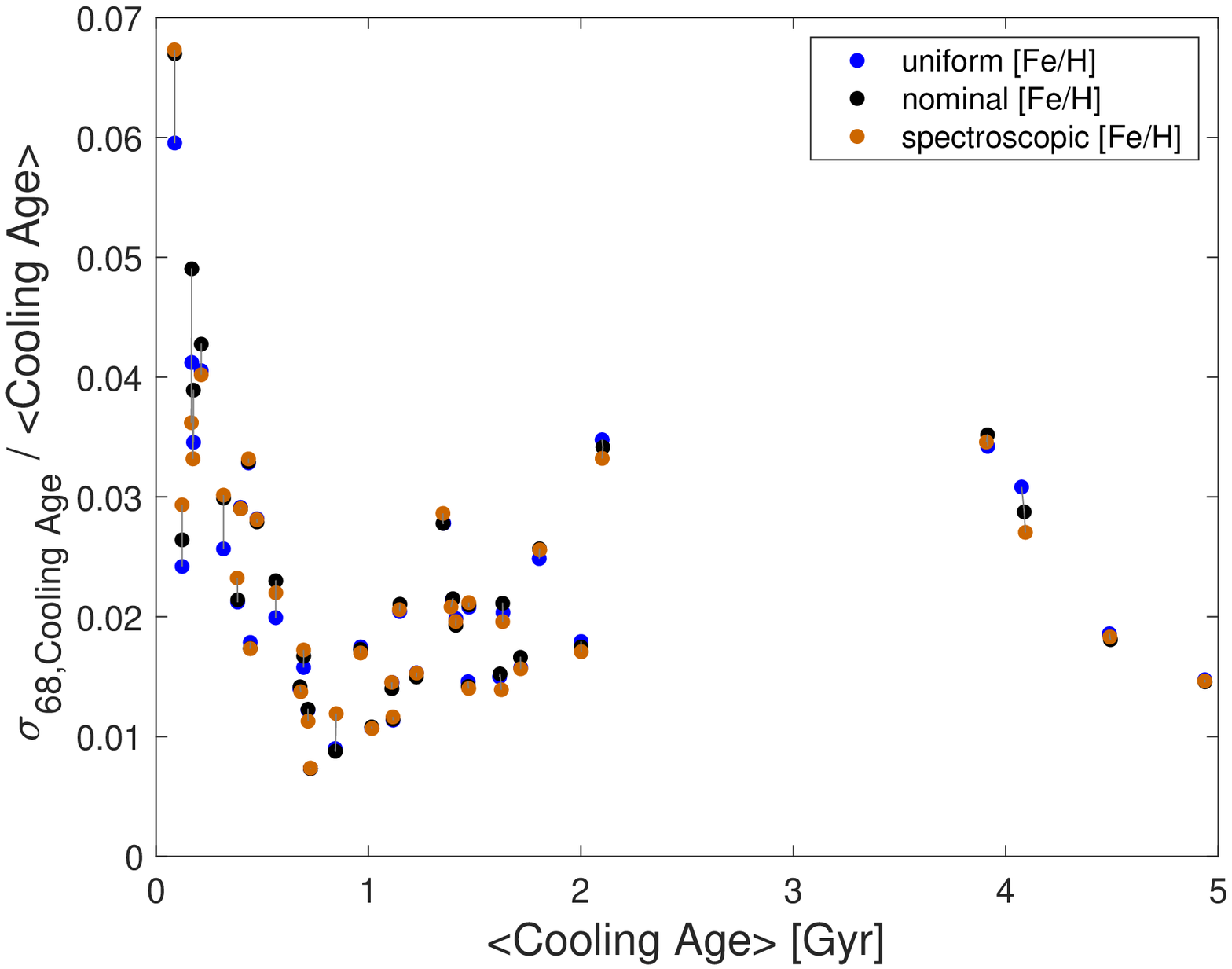}
  \hspace{1cm}
  \includegraphics[width=0.45\textwidth]{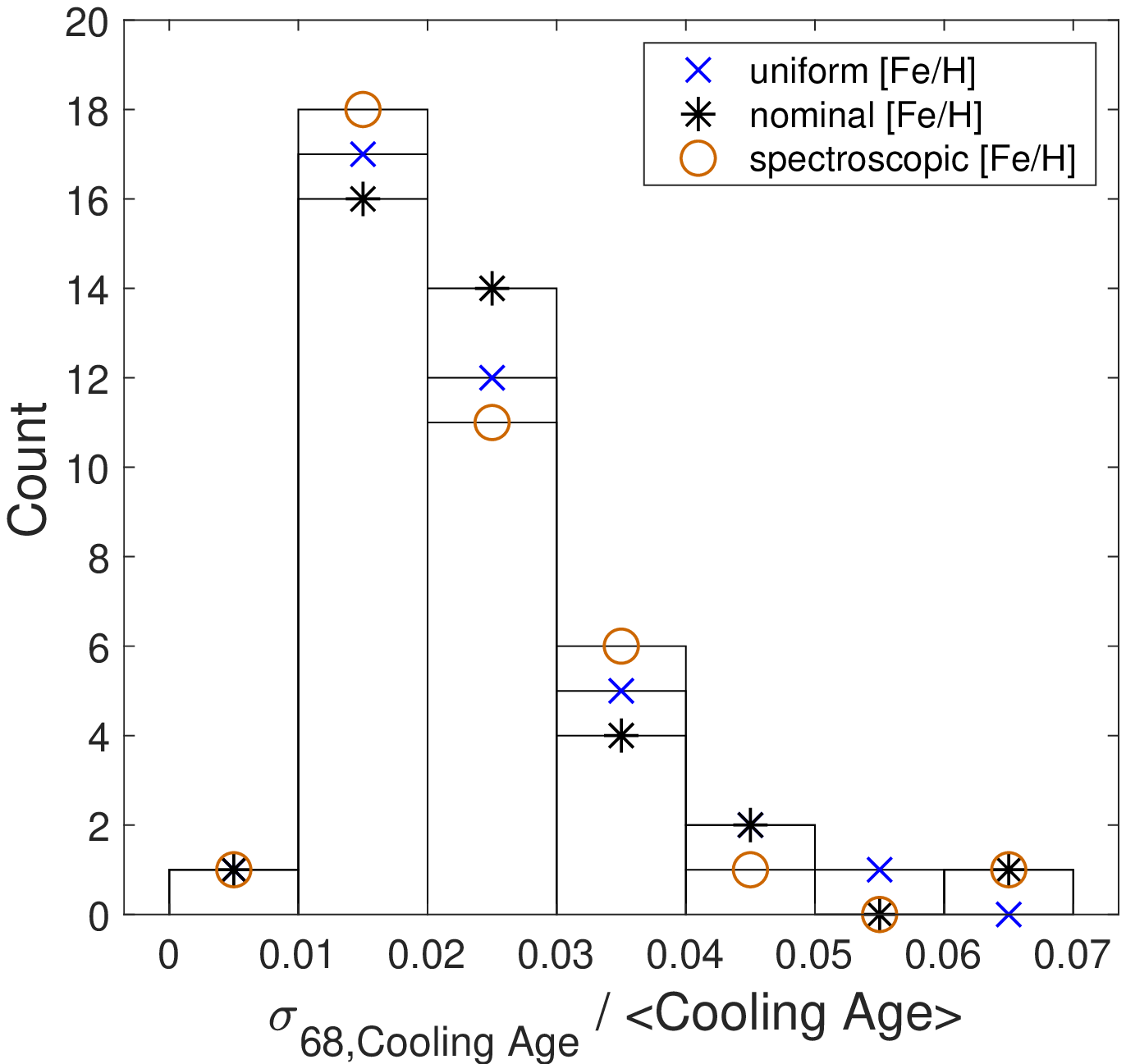}
  \caption{The WD cooling age uncertainties for the 38 WDs with spectroscopic metallicities (left panel). The three colors refer to posterior means with uniform (blue), nominal (black), and spectroscopic (orange) metallicity priors with an individual star's results connected by a black line. The binned uncertainties (right panel) demonstrate both high precision of these WD cooling ages and that the metallicity priors do not meaningfully impact cooling ages.}
\end{figure}

\begin{figure}  
    \vspace{0.5cm}
    \centering
    \includegraphics[width=0.65\textwidth]{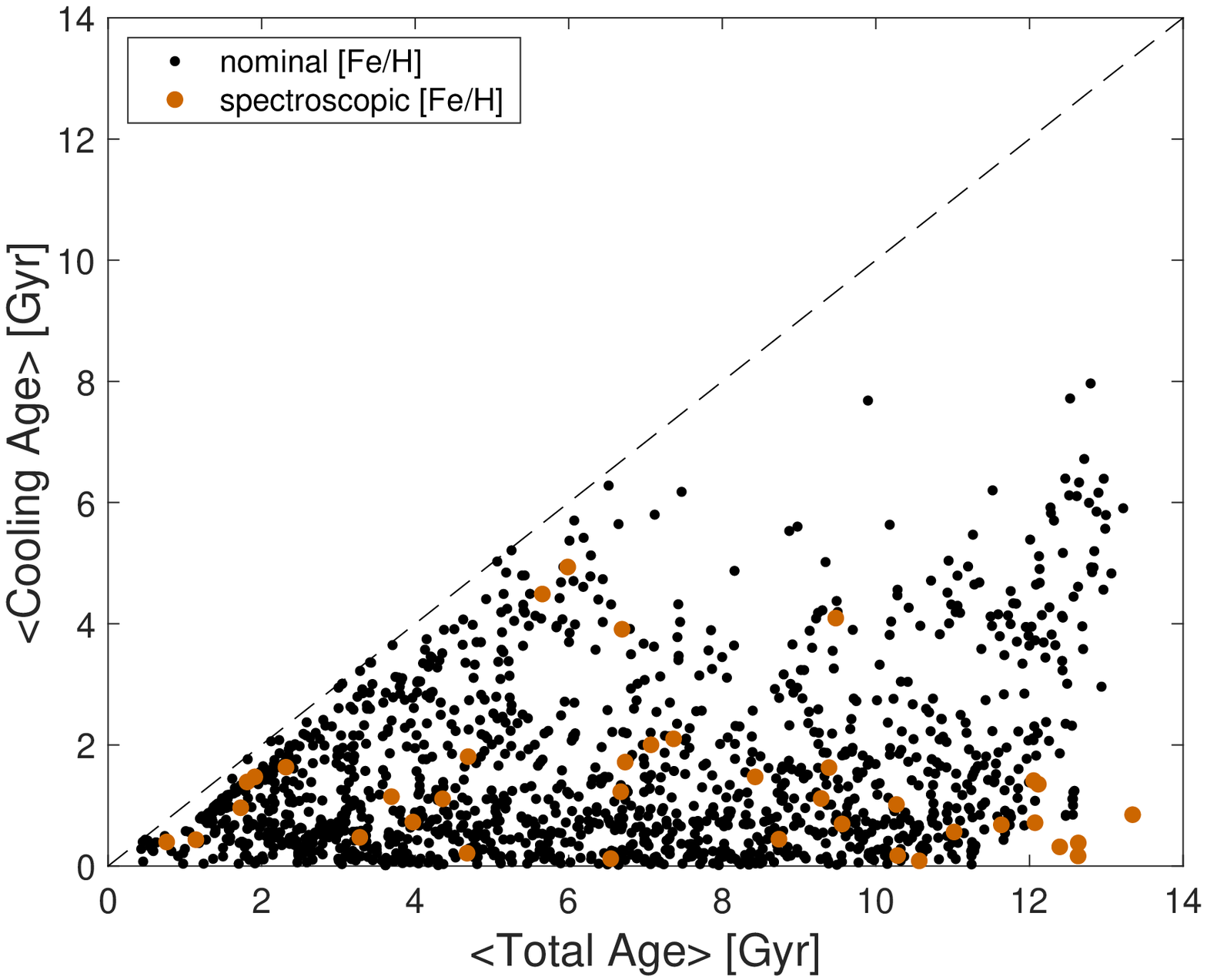} 
    \label{fig:my_label4}
    \caption{Posterior mean WD cooling age versus posterior mean total age for calculations performed under two metallicity priors.  Total age is always greater than the cooling age, creating the unit slope boundary indicated by the dashed line. WD temperatures and luminosities decrease with the vertical distance above the abscissa and WD precursor masses decrease to the right. The majority of these systems are 2 Gyr or older.}
\end{figure}


Figure 5 presents the posterior mean of the age (left panel) and of the mass (right panel) for each WD versus its fractional total age uncertainty.  The uncertainty in total age can be quite high for low mass stars.  Yet there are many stars with uncertainties in total ages of $\leq$ 0.20 and even quite a few with uncertainties $\leq$ 0.05.  The diagonal boundary of decreasing fractional uncertainties with increasing total ages is caused by the posterior distributions of stellar ages being constrained by the age of the Universe.  Older objects, which are necessarily younger than the Universe, have posterior distributions of age bound on the older side, decreasing their fractional uncertainty in total age.  

The left panel of Figure 6 presents the posterior means of total ages and their fractional uncertainties under each of the three  metallicity priors for the 38 stars with spectroscopic metallicities.  For all these stars, spectroscopic metallicity priors drive down the fractional total age uncertainty.   Specifically, averaging across the 38 stars, the fractional total age uncertainty incorporating the spectroscopic metallicity prior is 25.71\% lower than with the nominal metallicity prior and 29.04\% lower than with the uniform metallicity prior.  Also, intriguingly, the posterior means of age for most WDs are  somewhat higher with the spectroscopic metallicity prior.  The  spectroscopic metallicities of most stars are in the range of [Fe/H] = $-$0.2 to 0.0, whereas the uniform metallicity prior we chose was centered at [Fe/H] = $-$0.5.  Increased metallicity corresponds to more time on the main sequence and thus higher  total fitted ages.  Increased metallicity also mildly affects the posterior means of stellar masses as displayed in the right panel of Figure 6.

Figure 7 displays the posterior means of mass and total age for the 1372 systems, including the $\sigma_{\rm 68}$ uncertainties in both parameters. The low-mass WDs tend to have higher total age uncertainties (see Figure 5, right panel), as lower mass WDs evolved from lower mass precursors where a small change in precursor mass corresponds to a large change in both main sequence lifetime and total age. Thus small ZAMS mass uncertainties produce high total age uncertainties. Note that uncertainties in the IFMR are not incorporated here as this analysis was conducted with a single IFMR.  Conversely, the cooling age dominates the total age for old WDs above approximately two solar masses as these stars spent substantially less time on the main sequence.  Because the BASE-9 WD cooling model fits are highly precise, the total age uncertainties are low for objects that spend a large fraction of their lifespan as WDs. Figure 7 also demonstrates that high mass WDs with total ages greater than $\sim$6 Gyr are selected out of our sample.  These objects are harder to find for two reasons: high mass WDs are smaller than low mass WDs and therefore less luminous at a given effective temperature; they spend less time in the pre-WD evolutionary phases and therefore have cooled for much longer at a given total stellar age.  Both effects push most high mass WDs below Gaia's magnitude limit. \textbf{We note here the results of \citet{Bergeron2019}, who found that using Pan-STARRS photometry yielded a systematic offset compared to spectroscopically derived parameters, particularly for hotter WDs.  Specifically, for stars hotter than 12000 K, the photometrically-fit temperatures were low by $\sim$5\%, with this offset increasing to almost 10\% near 18000 K, at the upper temperature limit for our WDs with spectroscopic metallicities.  For these same stars the spectroscopic masses are $\sim$0.05 $M_\sun$ larger than those derived via photometry.  These offsets would naturally affect our total age and mass results, which in turn have an impact on the uncertainties. However, our purpose here is to study the internal age precision using high quality photometry and prallaxes along with spectroscopic metallicity priors, and not the external systematic uncertainties of the fit between the imperfect models and the data. Thus we did not incorporate this offset in our analysis.}


\begin{figure}  
\vspace{0.5cm}
\centering
  \includegraphics[width=0.45\textwidth]{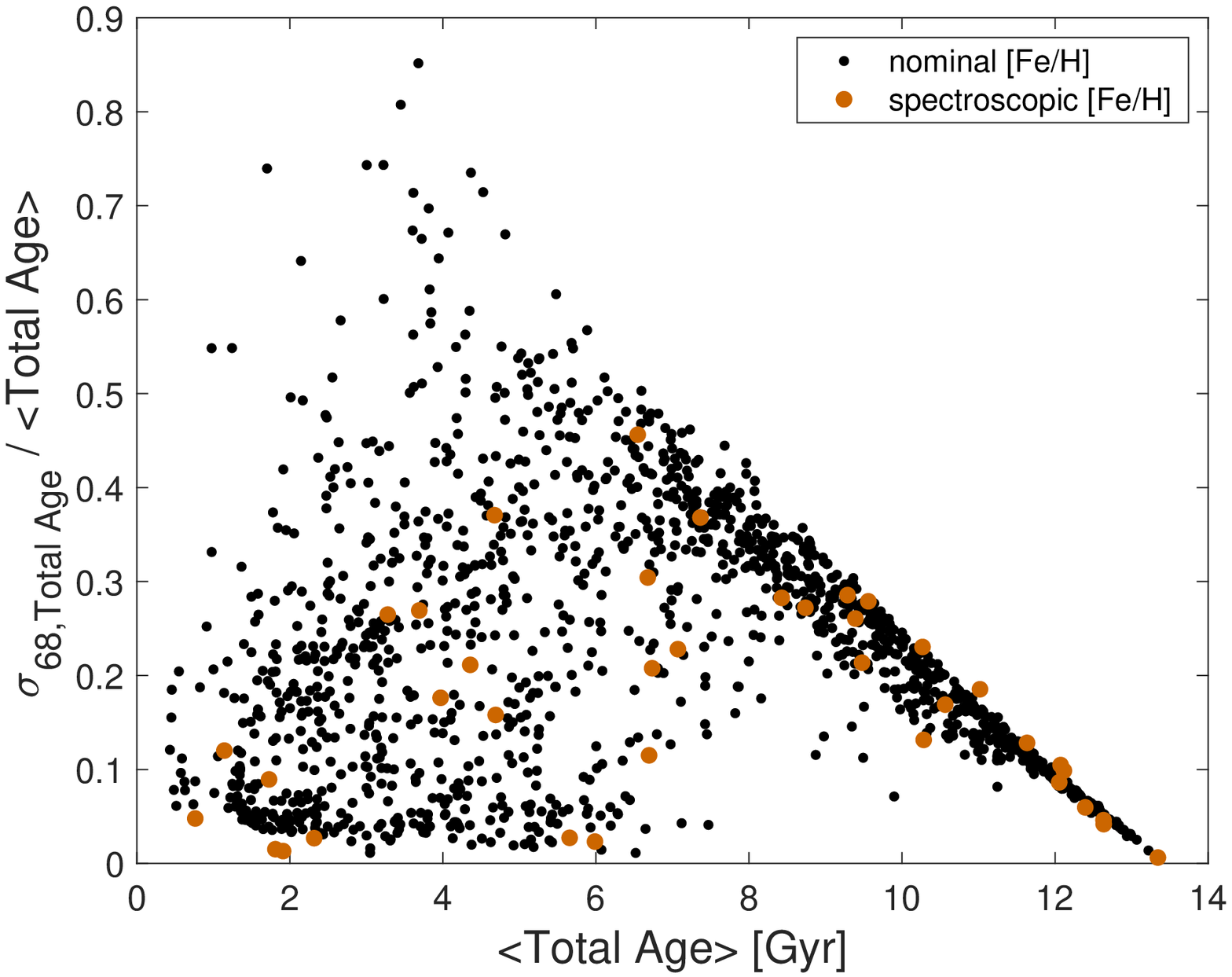}
  \hspace{1cm}
  \includegraphics[width=0.45\textwidth]{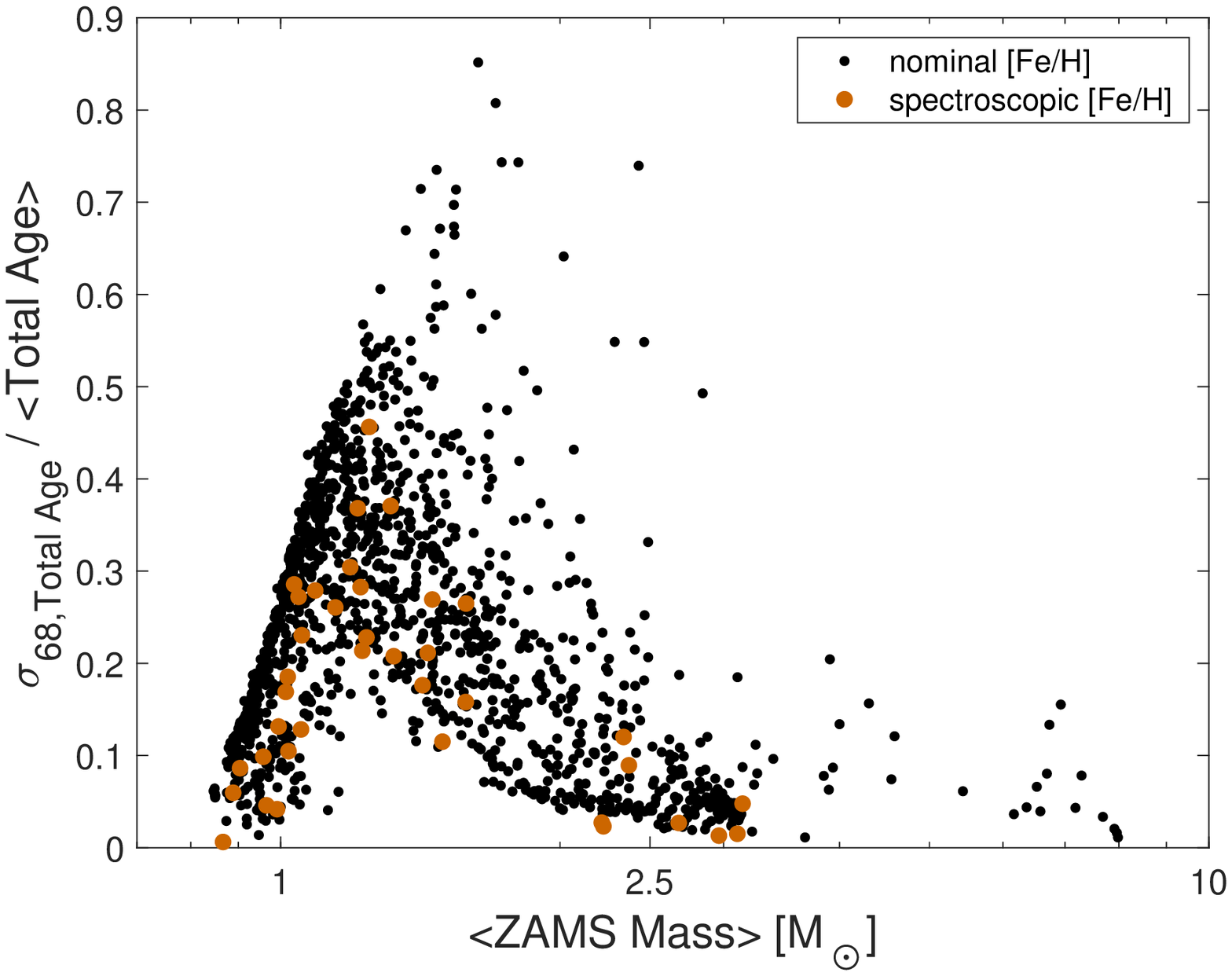}
  \caption{Posterior means of total age (left panel) and ZAMS mass (right panel) for the WDs vs. fractional uncertainties in total age for 1372 WD+MS binaries.  The majority of these systems have total ages $\leq$ 6 Gyr and fractional uncertainties in total ages $\leq$ 0.30, with many total age uncertainties $\leq$ 0.10.  WDs with total ages $>$ 8 Gyr follow a diagonal structure in the left panel because the posterior distributions of total stellar age are bounded above by the age of the Universe.  Many of these stars are likely not truly this old, but rather unresolved WD+WD binaries or the result of close binary star evolution.  In the right-hand panel there is an excluded zone of very high age uncertainty at low mass for the same reason --  these objects are the apparently oldest objects in our sample and so their posterior distributions are bound by the age of the Universe.}
\end{figure}

\begin{figure}  
  \centering
  \includegraphics[width=0.45\textwidth]{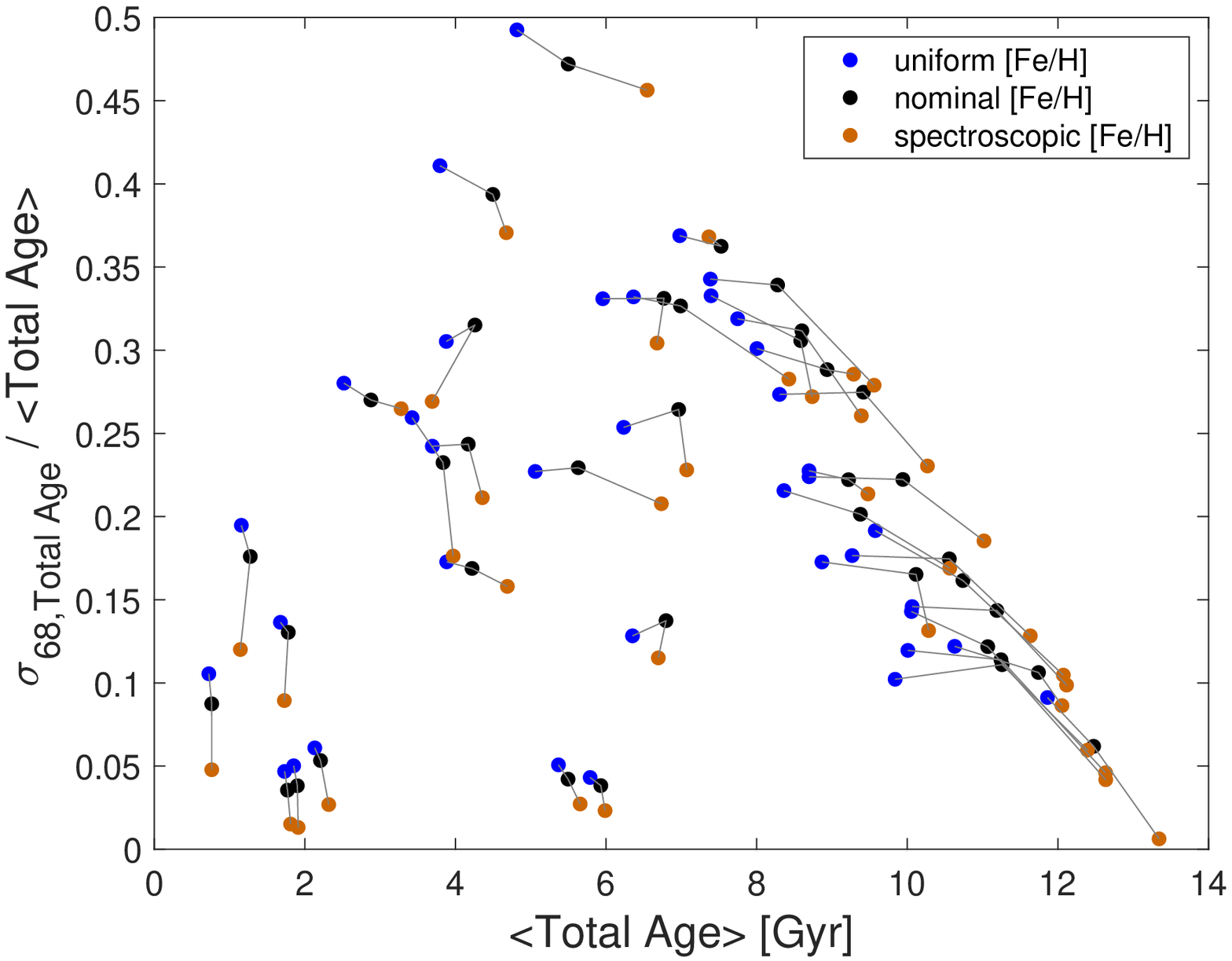}
  \hspace{1cm}
  \includegraphics[width=0.45\textwidth]{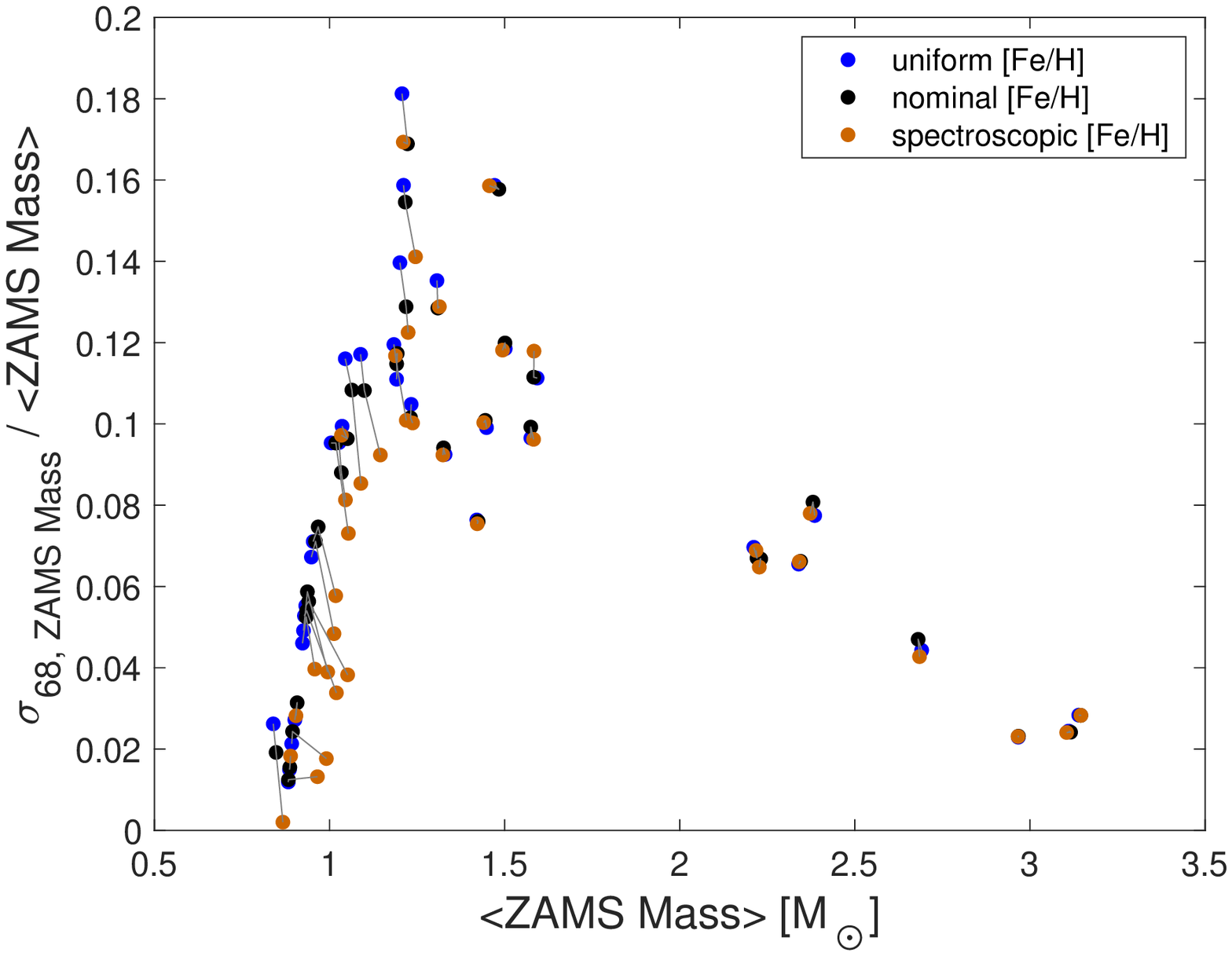}
  \caption{Uncertainty in total age vs. posterior mean of total age (left panel) and ZAMS mass uncertainty vs. posteror mean of zams mass (right panel) for the 38 WDs with spectroscopic metallicities from their MS companions.  BASE-9 was run three times for each of these systems with three different metallicity priors, demonstrating the degree to which metallicity information for the MS star constrains the total age for the WD, and thus the binary.  Individual stars run under the three metallicity priors are connected by lines.}
\end{figure}

\begin{figure}  
    \vspace{0.5cm}
    \centering
    \includegraphics[width=0.65\textwidth]{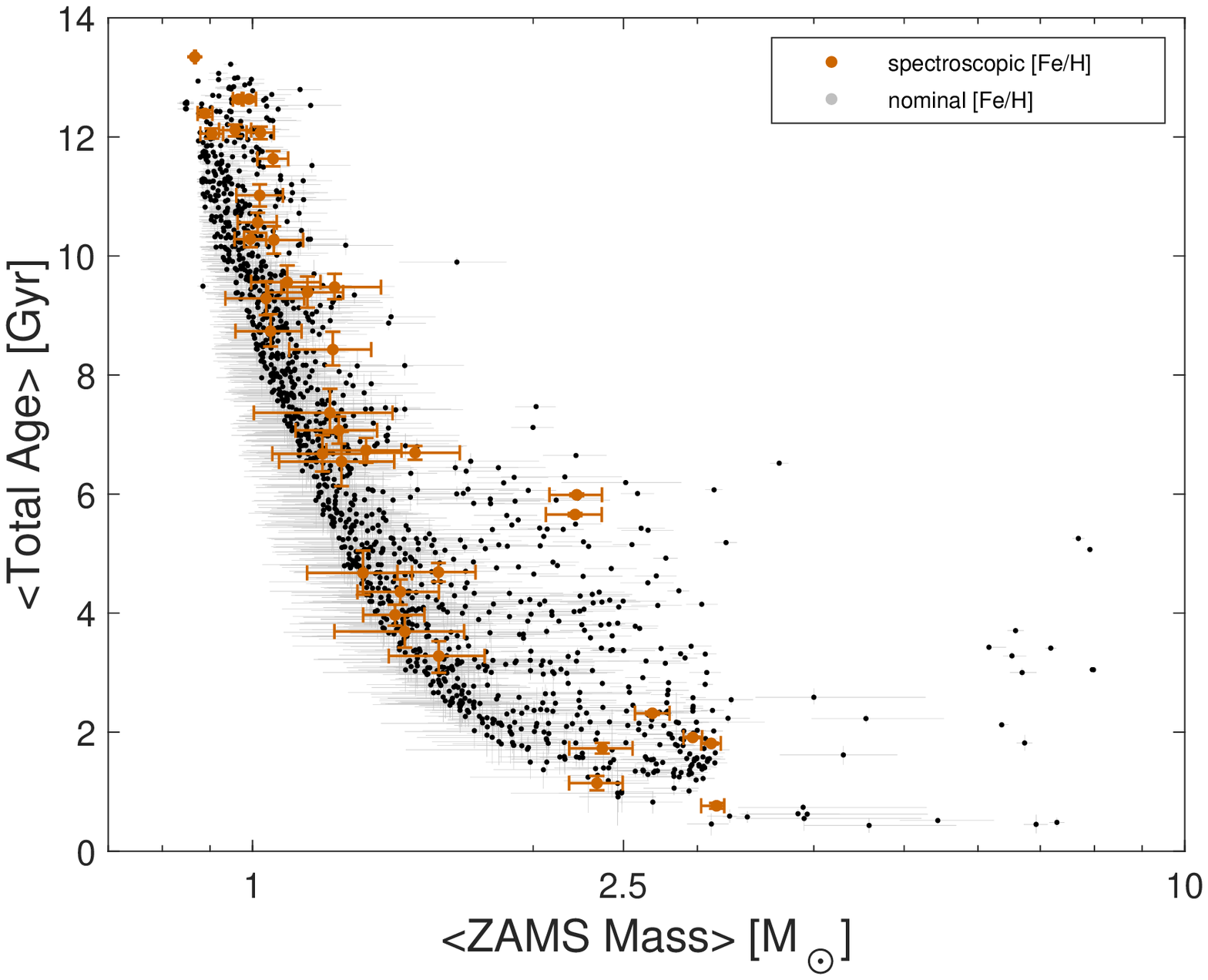}
    \caption{Estimated ZAMS mass and total age for all WDs in our sample. Total age uncertainties of less than 20\% are common for higher mass objects that spent the majority of their existence as WDs.  Systems with spectroscopic metallicities are indicated in orange whereas the rest of the sample are indicated in black.}
\end{figure}



\textbf{In order to study the effect of photometric and parallax precision on the resulting values of age precision, we selected 100 WDs that yielded fractional total age uncertainties $\leq$ 10\%, artificially degraded their photometry or parallax priors, and reran BASE-9 with these lower quality inputs. The photometric uncertainty for these 100 WDs in the five Pan-STARRS bands was typically 0.01 to 0.02 mag, though often somewhat worse in the $y$ and $z$ bands.  We increased photometric errors to a minimum of 0.02 mag in the first simulation and to a minimum of 0.05 mag in the second simulation.  We did not adjust the photometry itself, just the photometric uncertainties.  For the next three simulations, we kept the original photometric uncertainties, but degraded the parallax prior (fractional parallax uncertainty), which was often better than 1\%, to 1\%, 2\%, and 5\% in precision. Figure 8 shows the results of our simulations. For this sample, the original BASE-9 results yielded an average fractional total age uncertainty of 4.7\%.  Degrading the photometry to 0.02 or 0.05 mag uncertainties yielded average fractional total age uncertainties of 6.5\% and 15.2\%, respectively.  Degrading the parallax prior to 1\%, 2\%, and 5\% yielded average fractional total age uncertainties of 4.8\%, 5.8\%, and 11.0\%, respectively.  We conclude that typical $g, r, i$ photometric errors should be $\leq$ 0.02 ($z$ errors matter little if the other three bands are precise) and fractional parallax uncertainty should be $\leq$ 2\% in order to yield the best age precision.  Such high-quality data does not guarantee $\sim$5\% fractional total age uncertainties, however, because the total age of low mass WDs (with ZAMS mass $\leq$ 1.5 to 2 $M_\odot$) is dominated by the time these stars spent on the MS and therefore small uncertainties in the WD masses can propagate into large total age uncertainties.} 

\begin{figure}  
  \centering
  \includegraphics[width=0.45\textwidth]{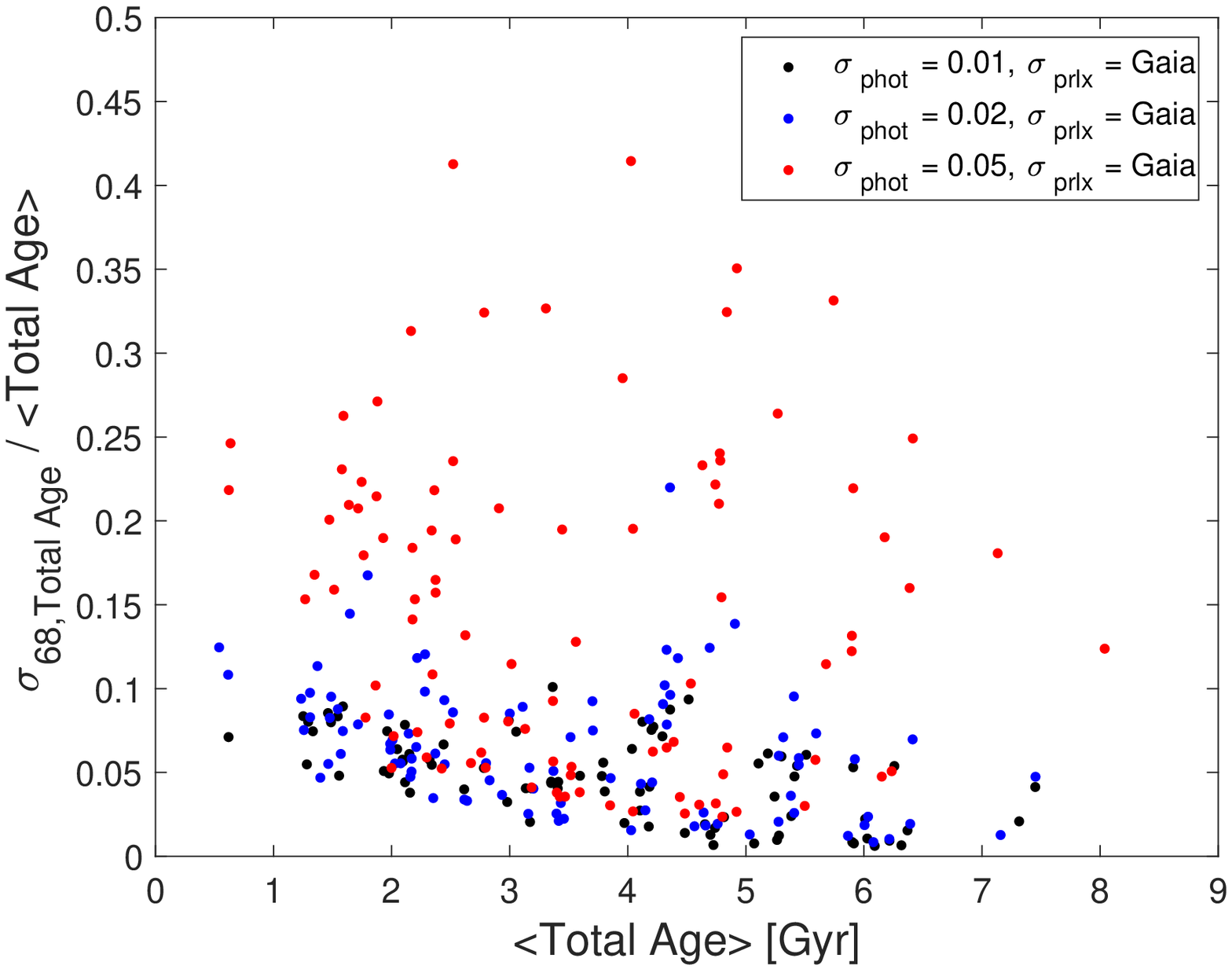}
  \hspace{1cm}
  \includegraphics[width=0.45\textwidth]{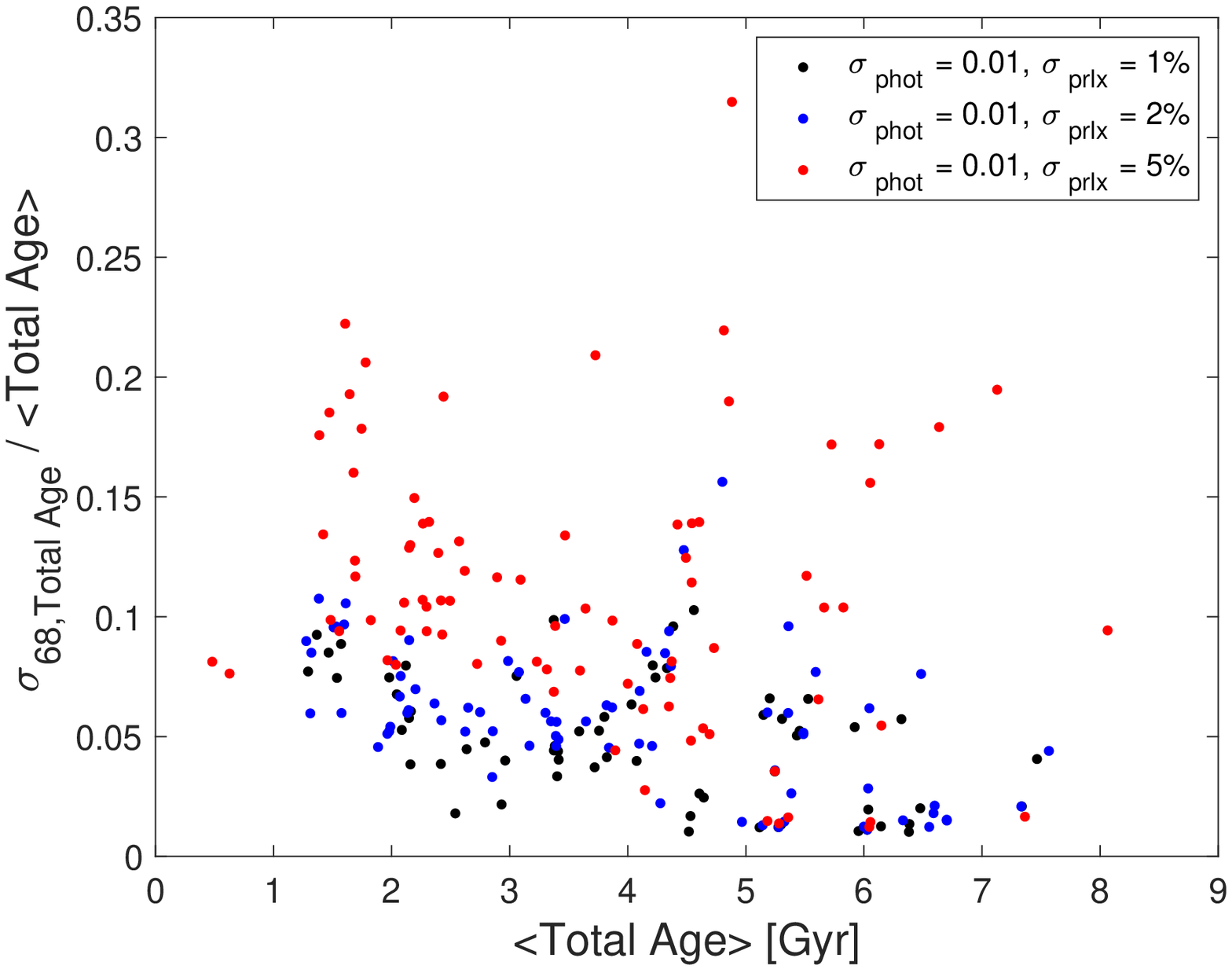}
  \caption{Resulting fractional total age uncertainties from increasing the Pan-STARRS photometric uncertainties (left panel) and Gaia parallax uncertainties (right panel) for our chosen 100 WDs. Black points represent the highest quality data in our simulations, with blue being slightly lower quality and red being the lowest quality. Lower quality data yields noticeably higher fractional total age uncertainties, with increases of more than 100\% from the best to worst case being quite common.}
\end{figure}

\section{Could Mass Transfer Impact Metallicity Estimates?}

We have studied the impact of spectroscopic metallicities from companion MS stars on the quality of total age estimates for WDs.  The two members of these WD+MS binary systems are assumed to be identical in metallicity.  Yet over their lifespans, the stars could interact and the MS star might accrete material from the evolving WD precursor, altering its surface composition.  Sufficient material transferred to the MS star could yield a MS  spectroscopic metallicity that is not characteristic of the WD during its evolution and potentially bias the total age estimate.  We can assess the likelihood of meaningful mass transfer onto the MS star by calculating the original separation between the stars within each pair.  The same mass loss that could potentially pollute the MS star also causes orbital evolution.  Using the \cite{Cummings2018} IFMR, we calculate the mass loss for each WD, and thereby the orbital expansion each system has  undergone.  We then use the present-day {\it projected} separations to derive the original stellar separations.  While 3-D separations are preferable, only 2\% of these pairs had sufficiently precise parallaxes to resolve the line-of-site separation of member stars at $\geq$ 3$\sigma$ using Gaia eDR3 data.  Because the projected stellar separations are a lower limit on the physical separations, this process is conservative for calculating whether the stars were close enough for meaningful mass transfer.  Figure 9 presents the distribution of the calculated original stellar projected separations.  The closest members within a binary were separated by an estimated 47\,AU prior to WD mass loss.  While there is evidence of contamination in two unrelated systems at moderate separations---HD 159062, with semi-major axis = 55 to 67 AU \citep{Bowler21} and HD 219634 = 51 Peg, with periastron at approximately 33 AU, \citep{Jorissen19}---most stars in our sample  were separated by $\geq$ 300 AU during their MS+MS evolution, and thus we expect the MS stars to be essentially uncontaminated by mass transfer.  Further evidence is that the 38 stars with spectroscopic metallicities show no apparent trend among higher metallicity stars to be nearer to their companions.  Therefore we expect a negligible impact of mass transfer on our analyses. 

\begin{figure}  
    \centering
    \includegraphics[width=0.65\textwidth]{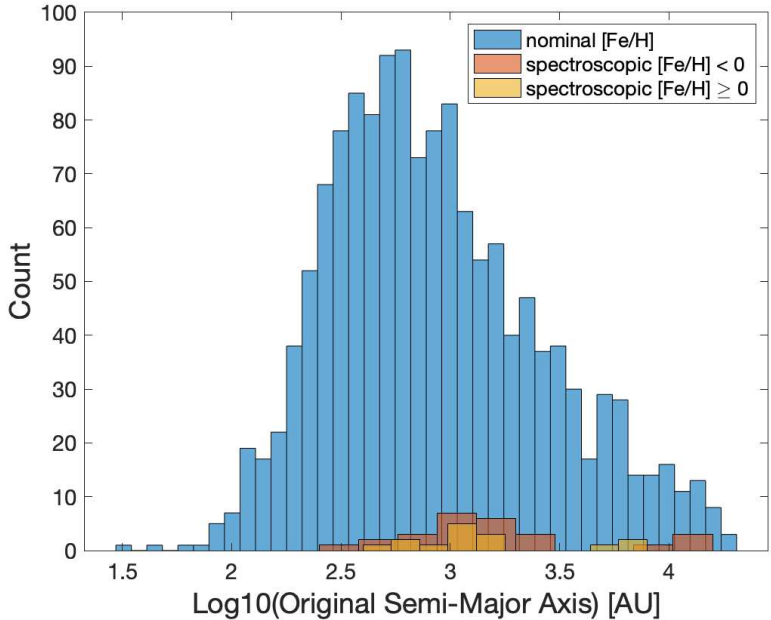}
    \caption{Projected separations of the WD-MS binaries during the MS+MS  phase.}
\end{figure}

\section{Conclusion}

In order to quantify the precision in WD cooling and total ages within a single adopted set of modern stellar evolution models, we combine a sample of WD+MS binaries identified within a Gaia DR2-based catalog with Gaia eDR3 parallaxes and Pan-STARRS photometry.  We select those systems with high-quality Gaia trigonometric parallaxes and precise Pan-STARRS photometry to create a sample of 1372 WD+MS pairs.  We further identify a subsample of 38 of these systems with spectroscopic metallicities from the APOGEE, LAMOST, or RAVE surveys.  We analyze all 1372 binary pairs with BASE-9, fitting modern stellar evolution and atmosphere models and thereby derive  posterior distributions for the cooling ages, total ages, metallicities, and ZAMS masses of the WDs. \textbf{We note that all of our analysis was conducted assuming each WD is a DA. While swapping atmospheres from a DA to a DB can lead to notable total age differences and will be investigated in the future, the purpose of this study is to showcase the improvement in total age precision when incorporating the spectroscopic metallicity prior. Additionally, the DA assumption should be correct for the majority of our cases based on known WD atmosphere populations.} Estimated WD cooling ages, irrespective of their metallicity priors, are often highly precise with fractional age uncertainties typically $\leq$ 5\% and often $\sim$2\%.  Total age estimates have poorer precision because they rely on timing estimates of the earlier phases of stellar evolution, yet the majority of these stars have total age uncertainties of $<$ 20\%, and many of the higher mass, older WDs have total age uncertainties substantially below 20\% (see Figure 5).  We compute the improvement in total age precision based on more informative metallicity priors for the 38 WDs for which we have spectroscopic metallicities.  We test three different priors: uniform (equal probability over the range [Fe/H] = $-$2.0 to +0.5), nominal (a Gaussian distribution centered at [Fe/H] = $-0.2$ with $\sigma$ = 0.5), and one that uses spectroscopic measurements (a Gaussian distribution centered on the measured value with a typical standard deviation of 0.065 dex).  Averaged across WDs, these three prior distributions yield fractional total age uncertainties over the 38 stars of 21.04\%, 20.19\%, and 16.77\%, respectively.  Higher mass WDs typically yield better total age precision, and for 8 WDs with spectroscopic metallicities and ZAMS masses $\geq$ 2.0 $M_\sun$, the mean total age uncertainties under the three priors are 8.06\%, 7.52\%, and 4.54\%.  We note that incorporating spectroscopic metallicities decreases the total age uncertainty substantially compared to the uniform metallicity prior, with the mean uncertainty dropping to 20\% of its value under the uniform prior for the 38-star sample and 47\% of its  value under the uniform prior for the 8 WDs with ZAMS masses $\geq$ 2.0 $M_\sun$. Because a spectroscopic metallicity prior helps refine the total age of a WD, where this information can be obtained, e.g. from a companion star in a binary or from another member of a star cluster or moving group, we recommend the additional observational effort.  For example, Gaia eDR3 contains approximately 16,000 WD+MS binaries \citep{Elbadry21} and we recommend spectroscopic surveys such as 4MOST \citep{deJong19},  SDSS-V \citep{SDSSV20}, or WEAVE \citep{Dalton12} follow-up these stars.  We finally note that the level of age precision (as good as 3.4\% for WDs with ZAMS masses $\geq$ 2.0 $M_\sun$) achieved here holds promise for chemical tagging studies where the MS stellar spectrum is studied in great detail to match it to other stars from the same formation episode.


\begin{sidewaystable}
  \begin{center}
     \begin{tabular}{|c|c|c|c|c|c|c|c|c|}
     \hline
        WD Gaia eDR3 ID & $\varpi$ [mas] & g & r & i & z & y & [Fe/H] & Total Age [Gyr] \\ [0.5ex]
        \hline
        3930951809593631360 & 8.166 (0.235) & 19.278 (0.010) & 19.092 (0.010) & 19.043 (0.010) & 19.083 (0.012) & 18.970 (0.016) & 0.171 (0.024) & 6.734 (0.208) \\
        147971725111014144 & 8.509 (0.282) & 19.182 (0.010) & 18.980 (0.010) & 18.911 (0.010) & 18.943 (0.018) & 18.909 (0.046) & 0.136 (0.103) & 4.689 (0.158) \\
        1877088553041925888 & 6.506 (0.135) & 18.045 (0.010) & 18.157 (0.010) & 18.277 (0.010) & 18.411 (0.020) & 18.404 (0.021) & -0.058 (0.042)  &    11.019 (0.185)\\
        1687736059280811008 & 7.374 (0.079) & 17.417 (0.010) & 17.544 (0.010) & 17.764 (0.010) & 17.911 (0.011) & 17.965 (0.054)    & -0.290 (0.100) &     8.737 (0.272)\\
        1554786751086765056 & 3.463 (0.521) & 20.693 (0.029) & 20.399 (0.016) & 20.275 (0.017) & 20.227 (0.041) & 19.882 (0.266) & -0.455 (0.089)     &     7.365 (0.368)\\
        1458024268941320064 & 10.591 (0.044) & 16.087 (0.013) & 16.338 (0.010) & 16.591 (0.010) & 16.832 (0.010) & 17.004 (0.010) & -0.180 (0.100)  &    4.675 (0.371)\\
        3491935581337073152 & 6.664 (0.070) & 17.037 (0.010) & 17.296 (0.012) & 17.541 (0.010) & 17.755 (0.010) & 17.717 (0.010) & 0.0418 (0.090)  &      12.633 (0.046)\\
        110863555565343360 & 7.323 (0.348) & 19.439 (0.020) & 19.081 (0.025) & 18.958 (0.033) & 19.063 (0.012) & 18.875 (0.044)  & -0.272 (0.032)    &      12.115 (0.099) \\
        968552852240732160 & 7.438 (0.211) & 19.170 (0.012) & 18.988 (0.014) & 18.951 (0.010) & 19.015 (0.010) & 18.855 (0.057) & 0.271 (0.031)     &      8.428 (0.283)\\
        679451907694651904 & 6.772 (0.106) & 17.560 (0.010) & 17.769 (0.010) & 17.981 (0.010) & 18.201 (0.016) & 18.301 (0.047) & -0.650 (0.100)  &     1.144 (0.120) \\
        667328516391473920 & 6.796 (0.257) & 19.002 (0.010) & 18.926 (0.020) & 18.920 (0.010) & 19.006 (0.010) & 19.000 (0.026) & -0.387 (0.062)     &      9.288 (0.286) \\
        724012189094256512 & 5.737 (0.194) & 18.718 (0.015) & 18.717 (0.010) & 18.844 (0.030) & 19.007 (0.038) & 18.846 (0.031)  & 0.254 (0.053)    &      11.634 (0.128)\\
        754878504143019136 & 7.841 (0.175) & 18.602 (0.010) & 18.518 (0.010) & 18.575 (0.010) & 18.643 (0.014) & 18.698 (0.028) & -0.060 (0.100)   &       10.268 (0.230) \\
         5457356901392996992 & 16.266 (0.185) & 19.007 (0.017) & 18.538 (0.010) & 18.377 (0.010) & 18.312 (0.011) & 18.255 (0.036) & 0.061 (0.080)  &  5.656 (0.027)\\
        763197172895543168 & 11.195 (0.317) & 20.009 (0.016) & 19.502 (0.026) & 19.308 (0.010) & 19.230 (0.011) & 19.221 (0.029)  & -0.132 (0.037)  &      5.987 (0.023)  \\
        767724270288678784 & 10.464 (0.145) & 18.302 (0.010) & 18.201 (0.011) & 18.242 (0.010) & 18.352 (0.010) & 18.332 (0.019) & 0.183 (0.040)    &       2.317 (0.027) \\
        1302430759290043648 & 7.035 (0.213) &  19.409 (0.011) & 19.159 (0.010) & 19.088 (0.010) & 19.126 (0.014) & 19.046 (0.031) & -0.660 (0.100) &   12.054 (0.086) \\
        802592670224001664 & 6.062 (0.338) & 19.555 (0.010) & 19.324 (0.010) & 19.296 (0.010) & 19.326 (0.024) & 19.218 (0.086) & 0.072 (0.030)     &     9.390 (0.261) \\
        6834975736126811520 & 9.966 (0.233) & 19.139 (0.012) & 18.848 (0.010) & 18.763 (0.010) & 18.783 (0.014) & 18.781 (0.021) & -0.218 (0.100)    &      7.071 (0.228) \\
        446203228969067392 & 10.246 (0.123 & 17.999 (0.010) & 17.997 (0.010) & 18.089 (0.011) & 18.206 (0.010) & 18.240 (0.020) & 0.039 (0.043)       &     1.811 (0.015) \\
        361054432874158848 & 6.112 (0.146) & 18.495 (0.010) & 18.512 (0.010) & 18.616 (0.010) & 18.742 (0.015) & 18.857 (0.028)  & 0.025 (0.089) &       9.560 (0.279) \\
        6309127501205488512 & 13.941 (0.098) & 17.450 (0.010) & 17.423 (0.010) & 17.496 (0.013) & 17.617 (0.010) & 17.618 (0.018) & -0.184 (0.100)  &     1.911 (0.013) \\
        1410345596469085184 & 6.712 (0.079) & 17.739 (0.018) & 17.889 (0.010) & 18.064 (0.010) & 18.273 (0.010) & 18.473 (0.025) & 0.010 (0.100)   &     3.279 (0.264)\\
        5705571754442134400 & 11.590 (0.045) & 15.516 (0.010) & 15.810 (0.010) & 16.091 (0.010) & 16.309 (0.010) & 16.478 (0.010) & 0.129 (0.090)    &     6.546 (0.456) \\
        6229228774355869824 & 11.796 (0.072) & 17.701 (0.010) & 17.113 (0.010) & 17.191 (0.010) & 17.313 (0.010) & 17.420 (0.015) & 0.132 (0.090)    &      12.072 (0.105) \\
        2116724500975907456 & 8.148 (0.106) & 18.560 (0.010) & 18.505 (0.010) & 18.537 (0.010) & 18.599 (0.010) & 18.667 (0.021) & -0.176 (0.028) &    4.357 (0.211) \\
        2334545730192266496 & 7.452 (0.197) & 18.822 (0.012) & 18.468 (0.022) & 18.377 (0.010) & 18.390 (0.010) & 18.270 (0.016) & -0.382 (0.080)      &      13.343 (0.006)  \\
        2576762266276447104 & 5.251 (0.145) & 18.105 (0.010) & 18.342 (0.010) & 18.606 (0.010) & 18.834 (0.016) & 19.018 (0.021) & -0.266 (0.039) &     0.763 (0.048)\\
        2346893104038050560 & 9.240 (0.409) & 20.242 (0.019) & 19.705 (0.011) & 19.469 (0.010) & 19.407 (0.049) & 19.302 (0.063) & -0.207 (0.090)   &      9.477 (0.214)  \\
        2367473247290858624 & 5.602 (0.314) & 19.263 (0.016) & 19.171 (0.013) & 19.238 (0.011) & 19.270 (0.030) & 19.334 (0.072) & -0.886 (0.170)  &      3.691 (0.269)  \\
        2442099751463686528 & 8.575 (0.070) & 16.066 (0.010) & 16.374 (0.010) & 16.668 (0.010) & 16.906 (0.010) & 17.055 (0.010) & -0.068 (0.090)       &      10.564 (0.169)  \\
        3540228743369046784 & 7.391 (0.079) & 17.163 (0.013) & 17.336 (0.010) & 17.509 (0.027) & 17.672 (0.037) & 17.687 (0.017)  & -0.434 (0.082) &      12.394 (0.060)  \\
        3185990694175594752 & 8.193 (0.447) & 20.262 (0.019) & 19.833 (0.015) & 19.617 (0.010) & 19.461 (0.017) & 19.542 (0.028) & -0.296 (0.090)    &      6.694 (0.115) \\
        1774286346148467968 & 8.407 (0.163) & 18.690 (0.010) & 18.591 (0.010) & 18.574 (0.010) & 18.647 (0.010) & 18.629 (0.030) & -0.353 (0.025)  &      6.677 (0.304) \\
        3963497529869428992 & 5.588 (0.091) & 17.312 (0.010) & 17.541 (0.010) & 17.800 (0.015) & 18.021 (0.016) & 18.001 (0.046) & -0.215 (0.021)   &      10.283 (0.132) \\
        3966668139152301568 & 5.849 (0.204) & 18.876 (0.011) & 18.293 (0.016) & 19.037 (0.010) & 19.136 (0.017) & 19.144 (0.046) & -0.411 (0.010)   &      1.728 (0.089)  \\
        3799009353404271488 & 20.739 (0.044) & 15.780 (0.010) & 15.864 (0.010) & 15.979 (0.010) & 16.127 (0.010) & 16.188 (0.010) & -0.186 (0.052)   &      3.970 (0.176)  \\
        3804750586512019840 & 6.661 (0.122) & 17.597 (0.012) & 17.703 (0.010) & 17.895 (0.010) & 18.070 (0.010) & 18.231 (0.023) & 0.117 (0.019)   &      12.636 (0.042)  \\
        \hline
     \end{tabular}
     \caption{General information for our 38 WDs where the spectroscopic [Fe/H] of the MS star is known.}
  \end{center}
\end{sidewaystable}

 
\acknowledgements

This material is based upon work supported by the National Science Foundation under Grant No.\ AST-1715718.  This work was supported in part by the DLR (German space agency) via grant 50\,QG\,1403.  David Stenning acknowledges the support of the Natural Sciences and Engineering Research Council of Canada (NSERC), RGPIN-2021-03985.  David van Dyk and David Stenning  acknowledge the support of a Marie-Skodowska-Curie RISE (H2020-MSCA-RISE-2019-873089) Grant provided by the European Commission.

This work has made use of data from the European Space Agency (ESA) mission {\it Gaia} (\url{https://www.cosmos.esa.int/gaia}), processed by the {\it Gaia} Data Processing and Analysis Consortium (DPAC, \url{https://www.cosmos.esa.int/web/gaia/dpac/consortium}). Funding for the DPAC has been provided by national institutions, in particular the institutions participating in the {\it Gaia} Multilateral Agreement.

The Pan-STARRS1 Surveys (PS1) and the PS1 public science archive have been made possible through contributions by the Institute for Astronomy, the University of Hawaii, the Pan-STARRS Project Office, the Max-Planck Society and its participating institutes, the Max Planck Institute for Astronomy, Heidelberg and the Max Planck Institute for Extraterrestrial Physics, Garching, The Johns Hopkins University, Durham University, the University of Edinburgh, the Queen's University Belfast, the Harvard-Smithsonian Center for Astrophysics, the Las Cumbres Observatory Global Telescope Network Incorporated, the National Central University of Taiwan, the Space Telescope Science Institute, the National Aeronautics and Space Administration under Grant No. NNX08AR22G issued through the Planetary Science Division of the NASA Science Mission Directorate, the National Science Foundation Grant No. AST-1238877, the University of Maryland, Eotvos Lorand University (ELTE), the Los Alamos National Laboratory, and the Gordon and Betty Moore Foundation.

Funding for the Sloan Digital Sky Survey IV has been provided by the Alfred P. Sloan Foundation, the U.S. Department of Energy Office of Science, and the Participating Institutions. SDSS-IV acknowledges support and resources from the Center for High Performance Computing  at the University of Utah. The SDSS website is www.sdss.org.  SDSS-IV is managed by the Astrophysical Research Consortium for the Participating Institutions of the SDSS Collaboration including the Brazilian Participation Group,  the Carnegie Institution for Science, Carnegie Mellon University, Center for Astrophysics | Harvard \& Smithsonian, the Chilean Participation Group, the French Participation Group, Instituto de Astrof\'isica de Canarias, The Johns Hopkins University, Kavli Institute for the Physics and Mathematics of the Universe (IPMU) / University of Tokyo, the Korean Participation Group, Lawrence Berkeley National Laboratory, Leibniz Institut f\"ur Astrophysik Potsdam (AIP),  Max-Planck-Institut f\"ur Astronomie (MPIA Heidelberg), Max-Planck-Institut f\"ur Astrophysik (MPA Garching), Max-Planck-Institut f\"ur Extraterrestrische Physik (MPE), National Astronomical Observatories of China, New Mexico State University, New York University, University of Notre Dame, Observat\'ario Nacional / MCTI, The Ohio State University, Pennsylvania State University, Shanghai Astronomical Observatory, United Kingdom Participation Group, Universidad Nacional Aut\'onoma de M\'exico, University of Arizona, University of Colorado Boulder, University of Oxford, University of Portsmouth, University of Utah, University of Virginia, University of Washington, University of Wisconsin, Vanderbilt University, and Yale University.

Guoshoujing Telescope (the Large Sky Area Multi-Object Fiber Spectroscopic Telescope LAMOST) is a National Major Scientific Project built by the Chinese Academy of Sciences. Funding for the project has been provided by the National Development and Reform Commission. LAMOST is operated and managed by the National Astronomical Observatories, Chinese Academy of Sciences.

Funding for RAVE has been provided by: the Leibniz Institute for Astrophysics Potsdam (AIP); the Australian Astronomical Observatory; the Australian National University; the Australian Research Council; the French National Research Agency; the German Research Foundation (SPP 1177 and SFB 881); the European Research Council (ERC-StG 240271 Galactica); the Istituto Nazionale di Astrofisica at Padova; The Johns Hopkins University; the National Science Foundation of the USA (AST-0908326); the W. M. Keck foundation; the Macquarie University; the Netherlands Research School for Astronomy; the Natural Sciences and Engineering Research Council of Canada; the Slovenian Research Agency; the Swiss National Science Foundation; the Science \& Technology Facilities Council of the UK; Opticon; Strasbourg Observatory; and the Universities of Basel, Groningen, Heidelberg and Sydney. This work has made use of data from the European Space Agency (ESA) mission Gaia (https://www.cosmos.esa.int/gaia), processed by the GaiaData Processing and Analysis Consortium (DPAC, https://www.cosmos.esa.int/web/gaia/dpac/consortium). Funding for the DPAC has been provided by national institutions, in particular the institutions participating in the Gaia Multilateral Agreement.

\bibliographystyle{mnras}   
\bibliography{WD_generic} 

\end{document}